\documentstyle[aps,epsfig]{revtex} \begin{document}

\draft
\title{   Scaling in dynamical Turing pattern formation:   \\
        density of defects frozen into permanent patterns  }

\date{ February 28, 2000 }

\author{ Jacek Dziarmaga }

\address{ Los Alamos National Laboratory, Theory Division T-6,
          MS-B288, Los Alamos, NM 87545, USA                     \\
          and M.Smoluchowski Institute of Physics, 
          Jagiellonian University, Krak\'ow, Poland              \\
          dziarmaga@t6-serv.lanl.gov                             }

\maketitle
\tighten

\begin{abstract}

{\bf We estimate density of defects frozen into a biological Turing
pattern which was turned on at a finite rate. A self-locking of gene
expression in individual cells, which makes the Turing transition
discontinuous, stabilizes the pattern together with its defects. A
defect-free pattern can be obtained by spatially inhomogeneous activation
of the genes. }

\end{abstract}

\subsection*{ Motivation and summary of results }

  Long time ago Turing pointed out \cite{turing} that simple
reaction-diffusion (RD) systems of equations can account for formation of
biological patterns.  The mainstream of research, as reviewed in
Ref.\cite{review}, is devoted to RD models in continuous space. The
continuum RD-patterns are smooth and nonpermanent. On the other hand, it
is an empirical fact that even the nearest neighbor cells can differ
sharply in their biological functions and their sets of expressed genes.
Moreover many biological patterns are permanent. Even the most primitive
viruses, like the much studied bacteriophage $\lambda$ \cite{ptashne},
possess genetic switches that discriminate between different developmental
pathways and make a once chosen pathway permanent. It is reasonable to
assume that cells of higher organisms can also lock their distinctive sets
of expressed genes.

  The Turing patterns on figures in the review \cite{review} are
contaminated with defects. If we insist on pattern permanence, we must
accept that patterns are permanent together with their defects. Sometimes,
like for the animal coat patterns, permanent defects can provide an animal
with its own characteristic life-long but not inheritable "fingerprints".
In other cases, like formation of vital organ structures, a single defect
can be fatal. In this situation it is important to understand better the
origin of defects.

  In this paper we use a simple toy model which in principle should give a
homogeneous Turing pattern. Defects are particularly manifest on such a
simple background. The model has two genes $A$ and $B$. The genes are
strong mutual repressors. The strong intracellular mutual inhibition is
the factor responsible for pattern permanence. Both genes are activated
simultaneously in a given cell when a level of their common activator $a$
exceeds its critical value $a_c$. Pattern formation in RD models of
Ref.\cite{review} was simulated with fixed model parameters. In this paper
we turn on the activator level $a$ at a finite rate to find a scaling
relation between density of defects and the rate. Strong mutual
intracellular inhibition stabilizes the pattern together with its defects.
We obtain permanent domains of $A$-phase and domains of $B$-phase divided
by sharp cell-size boundaries.

  We also show that an inhomogeneous activation of the genes can result in
a perfect defect-free homogeneous pattern. At first the activator $a$
exceeds $a_c$ in a small seed area where, say, the gene $A$ is chosen.
Then $a$ slowly spreads around gradually activating more and more cells.
The initial choice of $A$ is imposed via intercellular coupling on all
newly activated cells. The inhomogeneous activation can be sufficiently
characterized by a velocity $v$ with which the critical $a=a_c$ surface
spreads. Thanks to the strong mutual intracellular inhibition there is a
nonzero threshold velocity $v_c$, such that for $v<v_c$ the formation of
defects is completely suppressed. In this way the very mutual inhibition
which is responsible for stability of defects can be harnessed to get rid
of them.

  The genetic network that we use in our toy model is functionally
equivalent to the genetic toggle switch which was syntetized by the
authors of the recent paper \cite{nature}. In that paper the network is
studied experimentally in a single "cell". It would be interesting to
generalize the experiment to a "multicellular" structure.

\subsection*{ The toy model }

  For the sake of definiteness we take a genetic network with two genes
$A$ and $B$. $A$ and $B$ are mutual repressors. The network is symmetric
under exchange $A \leftrightarrow B$. Expression of both genes is
initiated by a common activator $a$. Let $A(t,\vec{x})$ and $B(t,\vec{x})$
denote time-dependent protein concentrations in the cell at the position
$\vec{x}$. $\vec{x}$ belongs to a discrete square lattice with a lattice
constant of $1$. Evolution of the protein concentrations is described by
the stochastic differential equations

\begin{eqnarray}\label{A,B}
&& \dot{A}(t,\vec{x}) \; = \; R \; S_A(t,\vec{x}) \; - \; A(t,\vec{x}) \;\;,\\
&& \dot{B}(t,\vec{x}) \; = \; R \; S_B(t,\vec{x}) \; - \; B(t,\vec{x}) \;\;.
\end{eqnarray}
The last terms in these equations are responsible for the protein
degradation. $R$ is a transcription rate. $S_{A,B}(t,\vec{x}) \in \{ 0,1
\} \;$ are dichotomic stochastic processes. They switch on $(0\rightarrow
1)$ and off $(1 \rightarrow 0)$ transcription of a given gene. For
simplicity the processes are assumed to have the same constant switch-off
rate $r^{\rm off}$. The switch-on rates depend on concentrations

\begin{eqnarray}\label{pA^on,pB^on}
&& r^{\rm on}_A(t,\vec{x}) \; = \; a(t) \; 
   F\left[ -W \; B(t,\vec{x}) + 
            V \; \sum_{{\rm n.n.}\vec{y}} A(t,\vec{y}) \right] \;\;,\\
&& r^{\rm on}_B(t,\vec{x}) \; = \; a(t) \; 
   F\left[ -W \; A(t,\vec{x}) + 
            V \; \sum_{{\rm n.n.}\vec{y}} B(t,\vec{y}) \right] \;\;.
\end{eqnarray}
$W,V$ are positive coupling constants, $a(t)$ is a concentration of the
activator. $F[z]$ is a smooth step-like sigmoidal function; the function
$F[z]=10^3 \exp(z-2.2)/[1+\exp(z-2.2)]$ was used in our numerical
simulations. In this model the genes A and B are mutual repressors
$(W>0)$. There is a "ferromagnetic" coupling between nearest-neighbor
cells $(V>0)$; expression of $A$ in a given cell enhances expression of
$A$ in its nearest neighbors.

 The model is motivated by a genetic switch between two mutual repressors
like the one studied in the phage $\lambda$ \cite{ptashne} and in the {\it
E. coli} switch \cite{nature}. The mutual repressors have a common
promoter site on DNA. A necessary condition for expression of any of them
is a binding of an activator molecule to their promoter site
\cite{koetal}. The concentrations $A$ and $B$ influence its affinity to
the promoter site. The gene expression is intermittent because of binding
and unbinding of activator molecules. The nearest-neighbor coupling is
possible thanks to signalling through intercellular membrane channels.
 
 In an adiabatic limit, when switching of $S_{A,B}$ is much faster than
protein expression and degradation, the processes $S_{A,B}$ can be
replaced by their time averages,

\begin{eqnarray}\label{Aadiab,Badiab}
&& \dot{A}(t,\vec{x}) \; = 
\frac{Ra(t)F\left[-W \; B(t,\vec{x}) +
                   V \; \sum_{{\rm n.n.}\vec{y}} A(t,\vec{y})\right]}
     {r^{\rm off}+
       a(t)F\left[-W \; B(t,\vec{x}) +
                   V \; \sum_{{\rm n.n.}\vec{y}} A(t,\vec{y})\right]}
- \; A(t,\vec{x})
\;\;,\\
&& \dot{B}(t,\vec{x}) \; = \;  
\frac{Ra(t)F\left[-W \; A(t,\vec{x}) +  
                   V \; \sum_{{\rm n.n.}\vec{y}} B(t,\vec{y})\right]}
     {r^{\rm off}+
       a(t)F\left[-W \; A(t,\vec{x}) +  
                   V \; \sum_{{\rm n.n.}\vec{y}} B(t,\vec{y})\right]}
- \; B(t,\vec{x})
\;\;.
\end{eqnarray}
Here we temporarily neglect any noise terms.

\subsection*{ Attractor structure }

  In a subspace of uniform configurations $A(t),B(t)$ these equations
simplify to the dynamical system

\begin{eqnarray}\label{ABunif}
&& \dot{A} \; =        
\frac{ Ra F\left[ -W \; B + 2dV \; A \right]}
     { r^{\rm off} + a F\left[-W \; B + 2dV \; A\right]}
- \; A
\;\;,\\
&& \dot{B} \; = \;
\frac{ Ra F\left[ -W \; A + 2dV \; B \right]}
     { r^{\rm off} + a F\left[-W \; A + 2dV \; B\right]}
- \; B
\;\;,
\end{eqnarray}
where $2d$ is the number of nearest neighbors in $d$ dimensions.

\begin{figure}[bt]
\centerline{\epsfxsize=8cm \epsfbox{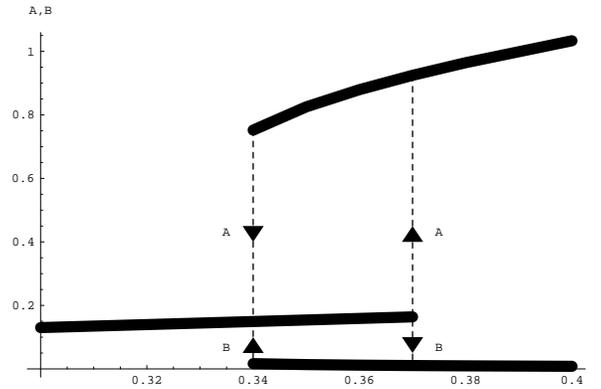}}
\vspace{-2cm}

\caption{ The thick lines are: $\alpha(a)$ (top), $\gamma(a)$ (middle),   
$\beta(a)$ (bottom). The vertical lines with arrows illustrate the
discontinuous jumps by the concentrations $A$ and $B$ during the 
$\alpha\beta\rightarrow\gamma\gamma$ transition at $a_{c_1}\approx 0.34$,
and the $\gamma\gamma\rightarrow\alpha\beta$ transition at $a_{c_2}\approx
0.37$. Model parameters used in this graph are: $R=4,W=3,V=1,d=2,r^{\rm   
off}=10^3$.}

\label{jumps}
\end{figure}

  The RHS's of these equations define a velocity field on the $A-B$ plane,
which is not a gradient field. The velocity field has attractor structure
which depends on the activator level $a$. There are two critical activator
levels $ a_{c_1} < a_{c_2} $. For $ a < a_{c_1} $ there is one attractor
at $[A,B]=[\gamma(a),\gamma(a)]$ with an increasing function $\gamma(a)$.
In the range $ a_{c_1} < a < a_{c_2} $ there are three attractors: the old
$[\gamma(a),\gamma(a)]$ plus a new symmetric pair of
$[\alpha(a),\beta(a)]$ and $[\beta(a),\alpha(a)]$ with $ \alpha(a) >
\beta(a) $. For $ a_{c_2} < a $ there remain only the two broken symmetry
attractors $[\alpha(a),\beta(a)]$ and $[\beta(a),\alpha(a)]$. The
functions $\alpha(a),\beta(a)$ and $\gamma(a)$ are plotted in
Fig.\ref{jumps}.

  If we start in the $[A,B]=[0,0]$ state and slowly increase $a$-level,
the system will stay in the $\gamma\gamma$-phase until we reach
$a=a_{c_2}$. At $a=a_{c_2}^+$ the system will roll into $\alpha\beta$ or
$\beta\alpha$-phase. On the other hand, if we start from $a_{c_2}<a$ with
the system in, say, $\alpha\beta$-phase, then we will have to decrease $a$
down to $a=a_{c_1}$, where $\alpha\beta$ becomes unstable towards the
symmetric $\gamma\gamma$-phase. The discontinuous jumps of the
concentrations are illustrated in Fig.\ref{jumps}. This hysteresis loop is
characteristic for first order phase transitions. In the adiabatic limit,
where fluctuations are small, there are no short cuts via bubble
nucleation. When $a_{c_{1}}$ ($a_{c_{2}}$) is approached from above
(below), the correlation length of small fluctuations around this uniform
state diverges like in a continuous phase transition. The critical regime
is narrow in the adiabatic limit so we can rely on the mean field
approximation.

\subsection*{ A finite rate Turing transition }

  Let us think again about starting from $[A,B]=[0,0]$ and continuously
increasing $a(t)$ above $a_{c_2}$. At $a_{c_2}^+$ the $\gamma\gamma$ state
becomes unstable and the system has to choose between the $\alpha\beta$
and $\beta\alpha$ attractors. If $a(t)$ is increased at a finite rate,
then there are finite correlated domains which make the choice
independently. Despite divergence of the correlation length at
$a_{c_2}^-$, the critical slowing down results in a certain finite
correlation length $\hat{\xi}$ "frozen" into the fluctuations. This scale
defines density of defects in the Turing pattern. This effect is well
known in cosmology and condensed matter physics as Kibble-Zurek scenario
\cite{kz}. In those contexts the defects disappear rapidly as a result of
phase ordering kinetics. We will see that in our gene network model the
defect pattern is permanent. This effect results from a combination of the
histeresis loop and the discreteness of the cell lattice.

  To be more quantitative we substitute $A(t,\vec{x})=\gamma(a(t))+\delta
A(t,\vec{x})$ and $B(t,\vec{x})=\gamma(a(t))+\delta B(t,\vec{x})$ into
Eqs.(\ref{Aadiab,Badiab}) and linearize them in $\delta A,\delta B$. The
linearized equations can be diagonalized by $\phi=\delta A-\delta B$ and
$\psi=\delta A+\delta B$. After Fourier transformation in space 

\begin{equation}
\phi(t,\vec{x})=\int d^dk\; \tilde{\phi}(t,\vec{k}) \;
                            e^{i\vec{k}\vec{x}}  
\end{equation}
they become

\begin{eqnarray}
&& \dot{\phi}(t,\vec{k})=R s_{\phi}(t,\vec{k})+
   \frac{ r^{\rm off} R a(t) F'_a}{ [ r^{\rm off} + a(t) F_a ]^2}
   \left[ W \; \phi(t,\vec{k}) + V \; e_{\vec{k}} \;
   \phi(t,\vec{k})\right]
   -\phi(t,\vec{k})                   \;\;, \label{phi}\\
&& \dot{\psi}(t,\vec{k})=R s_{\psi}(t,\vec{k})+
   \frac{ r^{\rm off} R a(t) F'_a}{ [ r^{\rm off} + a(t) F_a ]^2}
   \left[ -W \; \psi(t,\vec{k}) + V \; e_{\vec{k}} \;
   \psi(t,\vec{k})\right]
   -\psi(t,\vec{k})                   \;\;, \label{psi}
\end{eqnarray}
where $e_{\vec{k}}=2\sum_{i=1}^{d}\cos k_i$ in $d$ dimensions and we
skipped the tildas over Fourier transforms. $F'[z]=dF[z]/dz$ and we used
the shorthands $F_a^{(')}=F^{(')}[(-W+2dV)\gamma(a(t))]$. $Rs_{\phi,\psi}$
are noises which result from fluctuations in $RS_{A,B}$. In the adiabatic
limit they can be approximated by white noises (both in space and in time)
with small magnitude.

\begin{figure}[bt]
\centerline{\epsfxsize=8 cm \epsfbox{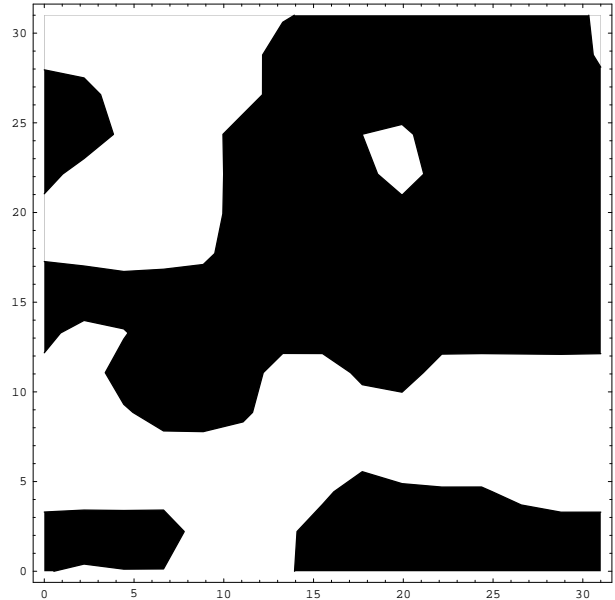}}
\vspace{-1cm}

\caption{ Permanent pattern obtained after switching-on the activator on a
$32 \times 32$ periodic lattice. It is a contour plot of $A-B$; white is 
$A$-rich ($\alpha\beta$) and black is $B$-rich ($\beta\alpha$). The
activator
was turned on as $a(t)=t/\tau$ with $\tau=32$ and $t\in (0,32)$. Model
parameters were the same as in Fig.\ref{jumps}. A discrete time step was  
$\Delta t=10^{-4}$. }

\label{patterns}
\end{figure}

\begin{figure}[bt]
\centerline{\epsfxsize=8 cm \epsfbox{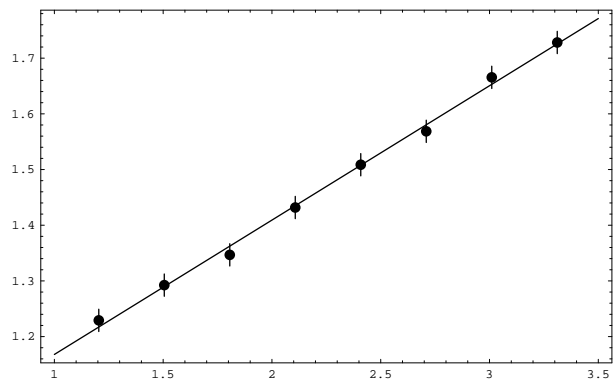}}
\vspace{-2cm}

\caption{ $\log(\hat{\xi})$ as a function of $\log(\tau)$. $\hat{\xi}$ was
obtained as an average domain size along a cross section through patterns 
like that in Fig.\ref{patterns}. For any given $\tau$ the average was
taken over outcomes of many simulations and over all the possible vertical
and horizontal cross sections. The vertical point size is a triple
standard deviation. The simulations were done on a $1024 \times 1024$    
lattice. The slope was fitted as $0.24 \pm 0.02$, which is consistent with
the predicted $0.25$.}

\label{logxi}
\end{figure}

  The next step is to linearize $a(t)$ around its critical value
$a(t)=a_{c_2}+t/\tau$, where $\tau$ is the transition rate. This
linearization gives

\begin{equation}
\frac{ r^{\rm off} R a(t) F'_a}{ [ r^{\rm off} + a(t) F_a ]^2}\;=\;
c_0 \;+\; c_1 \; \frac{t}{\tau} \;+\; O[(t/\tau)^2] \;.
\end{equation}
Approximating $e_{\vec{k}}=2d-\vec{k}^2$ in Eqs.(\ref{phi},\ref{psi}) and
keeping only leading terms in $t/\tau$ and in $k^2$ we get

\begin{eqnarray}
&& \dot{\phi}(t,\vec{k}) \;=\;
   R s_{\phi}(t,\vec{k}) \;+\;
   \left[ (\frac{c_1}{c_0})\frac{t}{\tau}-(c_0 V)\vec{k}^2 \right] \;
   \phi(t,\vec{k})                   \;\;,            \label{phi2}\\
&& \dot{\psi}(t,\vec{k}) \;=\;
   R s_{\psi}(t,\vec{k}) \;-\;
   [2c_0 W+c_0 V\vec{k}^2] \; \psi(t,\vec{k})  \;\;.  \label{psi2}
\end{eqnarray}
Here we used the identity $c_0[W+2dV]=1$, which has to be satisfied
because, by definition, $\phi(t,\vec{0})$ is a zero mode at $a_{c_2}$. The
$\psi$ modes are stable for any $\vec{k}$. The $\phi$-modes in the
neighborhood of $\vec{k}=\vec{0}$ become unstable for $t>0$ (or
$a_{c_2}<a$). Eq.(\ref{phi2}) is a standard linearized Landau model with
the symmetry breaking parameter $(c_1/c_0)(t/\tau)$ changing sign at
$t=0$. The length scale $\hat{\xi}$ frozen into fluctuations at $t>0$ can
be estimated following the classic argument given by Zurek \cite{kz}. For
$t<0$ the model (\ref{phi2}) has an instantaneous relaxation time $c_0\tau
/c_1|t|$ and an instantaneous correlation length $c_0\sqrt{V\tau
/c_1|t|}$. They both diverge at $t=0^-$. The fluctuations can no longer
follow the increasing $a(t)$ when their relaxation time becomes equal to
the time still remaining to the transition at $a=a_{c_2}$, $c_0\tau
/c_1|t| \approx |t|$. At this instant the correlation length is

\begin{equation}\label{hatxi}
\hat{\xi} \;\approx\; 
\left( \frac{V^{1/2}c_0^{3/4}}{c_1^{1/4}} \right) \;
\tau^{1/4} \;\;.
\end{equation}
This scale determines the typical size of the $\alpha\beta$- and
$\beta\alpha$-domains. The scaling relation $\hat{\xi}\sim\tau^{1/4}$ was
verified by numerical simulations illustrated at figures \ref{patterns}
and \ref{logxi}. The domain structures generated in the simulations turned
out to be permanent.

\begin{figure}[bt]
\centerline{\epsfxsize=8 cm \epsfbox{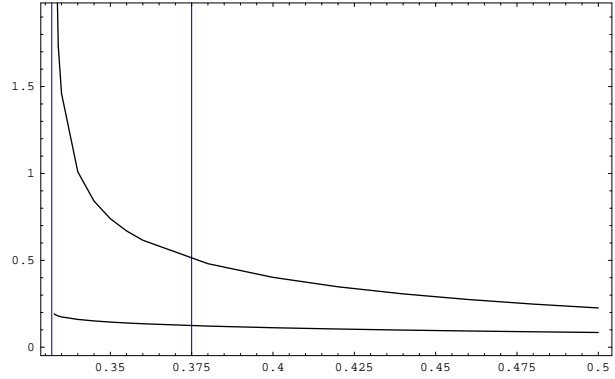}}
\vspace{-2cm}

\caption{ The correlation lengths of the fluctuations around the state
$\alpha\beta$ as functions of $a$. The vertical gridlines mark $a_{c_1}
\approx 0.332$ and $a_{c_2} \approx 0.375$. The larger correlation length  
diverges at $a_{c_1}$. These correlation lengths should be compared with
the lattice spacing which is $1$. The correlation lengths were obtained by
expanding $A(t,\vec{x})=\alpha(a)+\delta A(t,\vec{x})$ and
$B(t,\vec{x})=\beta(a)+\delta B(t,\vec{x})$, Fourier-transforming the
fluctuations in space and subsequent diagonalization for small $k$. }

\label{cl}
\end{figure}

  The domain structures are permanent because already at $a_{c_2}$ the
width of the domain wall interpolating between $\alpha\beta$ and
$\beta\alpha$ is less then the cell size (lattice spacing). The nearest
neighbor cells across the wall express different genes. The width (the
healing length) is determined by the longest length scale of fluctuations
around the $\alpha\beta$- or $\beta\alpha$-state. These correlation
lengths are plotted in Fig.\ref{cl}. For $a \geq a_{c_2}$ they are
substantially less than $1$. In the adiabatic limit, where the noises are
weak, the domain wall cannot evolve because it would have to overcome a
prohibitive potential barrier. On a cellular level the barrier originates
from the mutual inhibition between $A$ and $B$ in a single cell. Roughly
speaking, much above $a_{c_1}$ each cell is locked in its gene expression
state and insensitive to its nearest neighbors' states.

\subsection*{ Inhomogeneous activation }

  The intracellular mutual inhibition stabilizes the Turing pattern but it
also stabilizes the defects frozen into the pattern. With the
$\hat{\xi}\sim\tau^{1/4}$ scaling the number of defects is rather weakly
dependent on $\tau$. There may be not enough time during morphogenesis to
get rid of the defects by simply increasing $\tau$. However, it is
possible to generate a defect-free pattern by spatially inhomogeneous
switching of the activator level $a$. For example, its concentration can
exceed $a_{c_2}$ at one point at first, where the cells happen to pick (or
are forced to pick), say, $\alpha\beta$-phase, and then the activator can
gradually spread around so that the initial seed of $\alpha\beta$-cells
gradually imposes their choice on the whole system. For continuous
transitions this effect was described in Ref.(\cite{inhom}).

  The effect of defect suppression in inhomogeneous activation can be most
easily studied in a one dimensional version of the model (\ref{A,B}).
Suppose that a smooth activator front is moving across the one dimensional
chain of cells with a velocity $v$, $a(t,x)\approx a_{c_2}+ (vt-x)/v\tau$
close to $x=vt$ where $a=a_{c_2}$. For definiteness we impose two
asymptotic conditions: for $vt \ll x$ ( where $a<a_{c_2}$) the cells are
in the $\gamma\gamma$-state, and for $x \ll vt$ (where $a>a_{c_2}$) they
are in $\alpha\beta$-phase. We can expect that as the $a$-front moves to
the right it is followed by the $\alpha\beta$ front gradually entering the
area formerly occupied by the $\gamma\gamma$-phase. If the concentration
front is fast enough to move in step with the activator front, then the
$\alpha\beta$-phase will gradually fill the whole system. If, on the other
hand, the concentration front is slower than the activator front then the
front of the $\alpha\beta$-phase will lag behind the $a=a_{c_2}$ front.
The gap between the two fronts will grow with time. The gap will be filled
with the unstable $\gamma\gamma$-phase ($a>a_{c_2}$ behind the $a$-front).
When the gap becomes wide enough, then $\gamma\gamma$-state will be able
to decay towards the $\beta\alpha$-state. A domain of $\beta\alpha$-phase
will eventually be nucleated behind the $a$-front.  Now the
$\beta\alpha$-domain will grow behind the $a$-front until its front lags
sufficiently behind so that a new domain of $\alpha\beta$-phase will be
nucleated. In this way the activator front will leave behind a landscape
of alternating $\alpha\beta$- and $\beta\alpha$-domains qualitatively the
same as for homogeneous activation.

\begin{figure}[bt]
\centerline{\epsfxsize=8 cm \epsfbox{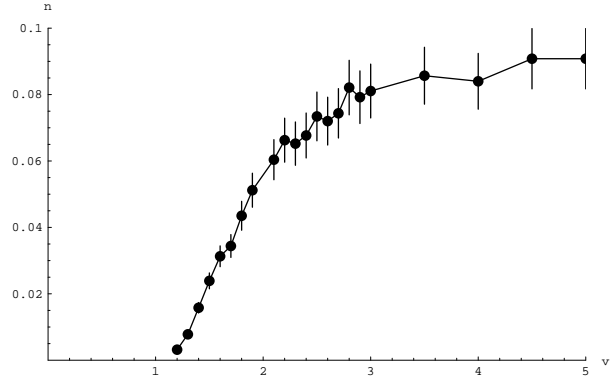}}
\vspace{-2cm}

\caption{ Density $n$ of domain walls between $\alpha\beta$ and
$\beta\alpha$-states behind an activator front with velocity $v$. The
activator was $a(t,x)=(vt-x)/v\tau$ for $x<vt$ and $a(t,x)=0$ for $vt<x$.
$1/v\tau=0.1$ was kept fixed so that the slope of $a$ versus $x$ was
independent of $v$. The model parameters were the same as in
Figs.\ref{jumps},\ref{patterns} but with $d=1$ instead of $2$ and $V=2$
instead of $1$ ($Vd=2$ as before). For these model parameters $v_c\approx
0.9$ in consistency with the numerical results. }

\label{front}
\end{figure}

  The success of the inhomogeneous activation depends on the relation
between the velocity $v$ of the $a$-front and that of the concentration
front. As illustrated in Fig.\ref{cl} fluctuations around the
$\alpha\beta$-state have two families of modes each with a different
correlation length. For any $a$ each $\vec{k}$-mode within each family has
a different diffusion velocity: a ratio of its wavelength to its
relaxation time. The lowest of these diffusion velocities, $v_c(a)$, is
the maximal velocity at which the $\alpha\beta$-phase can spread into the
area occupied by the $\gamma\gamma$-phase.  $v_c(a_{c_2})\equiv v_{c_2} >
0$ because at $a=a_{c_2}$ the $\alpha\beta$-state is stable (the
hysteresis loop again!). $v_c(a)$ increases with an increasing $a$. If
$v<v_{c_2}$ the $\alpha\beta$-front moves in step with the $a$-front; its
tail spreads into the $vt<x$ area imposing an $\alpha\beta$-bias on the
fluctuations around $\gamma\gamma$-state. The $\alpha\beta$-phase spreads
without nucleation of any $\beta\alpha$-domains. For $v<v_{c_2}$ a
defect-free uniform Turing pattern forms behind the activator front.
Results from numerical simulations of the inhomogeneous activation are
presented in Fig.\ref{front}.

\subsection*{ More complicated patterns }

  Finally, it is time to comment on more complicated models which are
expected to give more complicated patterns than the (in principle) uniform
pattern discussed so far. Let us pick a zebra pattern for example. For the
uniform pattern the first mode to become unstable in Eq.(\ref{phi2}) is
the $\vec{k}=\vec{0}$ mode. The final pattern has an admixture of
$\vec{k}$'s in a range $\approx \hat{\xi}^{-1}$ around $\vec{k}=\vec{0}$.
In distinction, for the zebra pattern the first unstable modes are those
on the circle $|\vec{k}|=2\pi/L$, where $L$ is the spacing between zebra
stripes. The final pattern has an admixture of $\vec{k}$'s in a ring of
thickness $\approx \hat{\xi}^{-1}$ around the circle $|\vec{k}|=2\pi/L$,
compare results for Swift-Hohenberg equation in Ref.\cite{lythe}. This
admixture results in defects frozen into zebra pattern. The inhomogeneous
activation can be applied in the zebra case too. In addition it can be
used to arrange the stripes. An activator spreading from an initial point
would result (at least close to the initial point) in concentric black and
white rings. A front of activator moving through the system would comb the
stripes perpendicular to the front.

{\bf Acknowledgements.} I would like to thank M.Sadzikowski and W.Zurek
for useful comments on the manuscript.

\end{document}